\documentclass[10pt,letter]{article}
  \pdfoutput=1
\newcommand{\shortversion}[1]{}
\newcommand{\longversion}[1]{#1}
\shortversion{\documentclass{llncs}}
\usepackage{url}
\usepackage{graphicx}
\usepackage{tikz}
\usepackage{setspace}
\usepackage{csquotes}

\usepgflibrary{shapes.geometric}
\usetikzlibrary{shapes.geometric,shapes.misc,shapes}
\usetikzlibrary{arrows}
\usetikzlibrary{shadows}
\usetikzlibrary{fadings}
\usetikzlibrary{positioning}
\usetikzlibrary{patterns}
\tikzset{vertex/.style={minimum height=5mm,minimum width=6mm,
      very thick,draw=black!50,
      top color=white,bottom color=black!20,inner sep=1.0pt}}
\tikzset{vertex2/.style={minimum height=5mm,minimum width=6mm,
      top color=white,bottom color=white!20,inner sep=1.0pt}}

\newcommand{\nop}[1]{}
\newcommand{\mybibitem}[1]{\bibitem[\arabic{bctr}]{}\stepcounter{bctr}}

\renewcommand{\refname}{
}

\newcounter{bctr}
\usepackage{hyperref}

\newcounter{ct_todo}
\usepackage{xcolor}

\usepackage[american]{babel}
\usepackage{microtype}

\usepackage{xspace}
\def\etal{et~al.\xspace}

\title{Answer Set Solving exploiting Treewidth and its Limits%
\thanks{This work has been supported by Austrian Science Fund (FWF): Y698. 
     The author is also affiliated with the University of
     Potsdam,~Germany.}
}

\author{Markus Hecher\longversion{\\[3pt]TU Wien, Austria, \href{mailto:hecher@dbai.tuwien.ac.at}{hecher@dbai.tuwien.ac.at}}}
\shortversion{\institute{Institute of Logic and Computation, TU Wien, Austria, \href{mailto:hecher@dbai.tuwien.ac.at}{hecher@dbai.tuwien.ac.at}\\
Keywords: Tree Decompositions, Dynamic Programming, Fixed-Parameter Tractability, Answer-Set Programming, Parameterized Complexity Theory}}

\begin{document}
\pagenumbering{arabic}

\maketitle


\pagenumbering{arabic}
\begin{abstract}%
Parameterized algorithms have been subject to extensive research of recent years
and allow to solve hard problems by exploiting a parameter
of the corresponding problem instances. 
There, one goal is to devise algorithms, where the runtime is exponential 
exclusively in this parameter. 
One particular well-studied structural parameter is treewidth.
Typically, a parameterized algorithm utilizing treewidth takes or computes a tree 
decomposition, which is an arrangement of a graph into a tree, and 
evaluates the problem in parts by dynamic programming on the tree decomposition.
In our research, we want to exploit treewidth in the context of
Answer Set Programming (ASP), a declarative modeling and solving framework, 
which has been successfully applied in several 
application domains and industries for years. 
So far, we presented algorithms for ASP 
for the full ASP-Core-2 syntax, which is competitive especially when it comes to counting answer sets. 
Since dynamic programming on tree decomposition lands itself well to 
counting, we designed a framework for projected model counting, which applies to ASP,
abstract argumentation and even to problems higher in the polynomial hierarchy.
Given standard assumptions in computational complexity, we established a novel methodology for showing lower bounds, and we showed that most worst-case runtimes of our algorithms cannot be significantly improved.
\end{abstract}
\section{Introduction}
Parameterized algorithms~\cite{CyganEtAl15} have
attracted considerable interest in recent years and allow to solve
hard combinatorial problems by utilizing a certain parameter of the problem instance.
Of particular interest is to devise algorithms, where
the runtime is polynomial and additionally depends on some computable function in the parameter. 
One structural and extensively studied
parameter is treewidth~\cite{BodlaenderKoster08,RobertsonSeymour86}.
Intuitively, treewidth measures the closeness of a graph to a tree based
on the observation that problems on trees are often easier than on
arbitrary graphs. 
A (parameterized) algorithm utilizing treewidth typically solves problem instances by \emph{dynamic programming (DP)}.
Thereby it takes a \emph{tree decomposition}, which is an arrangement of a graph (representation) of the given problem instance into a tree, and then evaluates the problem in parts.
Such dynamic programming algorithms are then sensitive to treewidth, which provides an upper bound on the worst-case runtime needed for each tree decomposition node of an (optimal) tree decomposition. 
However, in practice, solvers based on this idea can produce results for certain problems~\cite{CharwatWoltran19,LonsingEgly18} up to treewidth~80.
Further, there are observations that instances relevant for certain applications and different problems sometimes have small treewidth.
One particular problem, for which dynamic programming algorithms on tree decompositions were investigated~\cite{JaklPichlerWoltran09} is
Answer Set Programming.
\emph{Answer Set Programming (ASP)}~\cite{BrewkaEiterTruszczynski11} is a logic-based
declarative modeling language and problem solving framework 
where selected models, the \emph{answer sets}, of a given ASP program directly represent the solutions to the modeled problem. 
ASP has been applied
in several application domains, which accelerates 
the search for alternative solving methods. 
This raises then the question whether exploiting structural
parameters, e.g., treewidth, improves the performance of evaluation of ASP programs, which forms the topic of the thesis.
Jakl~\etal~\cite{JaklPichlerWoltran09} established a DP algorithm for disjunctive ASP that is linear in the size of the (ground) ASP program, but double exponential in
the treewidth of a certain graph representing of the program.
There is an implementation called~\emph{DynASP solver}~\cite{MorakEtAl11} that adheres to this idea.
In our work~\cite{FichteEtAl17a,FichteEtAl17b}, we presented an evaluation of extended ASP programs based on the full 
ASP-Core-2~\cite{aspcore2} syntax. 
Further, we also showed that the double exponential runtime in the treewidth can not be avoided  for ASP in general, unless the widely believed exponential time hypothesis (ETH) fails. This result is based on our novel methodology~\cite{FichteHecherPfandler19} for showing such lower bounds for problems even higher in the polynomial hierarchy.
Currently, we are investigating \emph{hybrid parameterized solving} techniques for ASP, where we aim to improve solving by a combination of both monolithic and parameterized solvers.
\section{Background}

\noindent\emph{Tree Decompositions.} Tree decompositions 
are built for a given graph and are a tree-like representation of the graph.
These tree decompositions are trees consisting of nodes,
where each node contains certain vertices of the given graph.
The so-called \emph{width} of a given tree decomposition 
corresponds to the largest number of vertices 
that is contained in one node (minus one).
The parameter \emph{treewidth} captures the ``tree-likeness''
of the given graph and corresponds to the smallest width
among all tree decompositions for the graph.
The treewidth intuitively reveals, how hard the given graph is 
when solving a certain problem. 
The smaller the treewidth of a given graph,
the more ``tree-like'' the graph and thus typically easier
to solve problems on the graph.
In particular, the treewidth of a graph that is a tree corresponds to one.
%

\noindent\emph{Dynamic Programming.}
Tree decompositions allow
to tackle hard problems by evaluating a certain problem-dependent graph representation of a problem instance
in parts, thereby being sensitive to the treewidth of the instance.
This evaluation is done by means of {dynamic programming (on the tree decomposition)},
where the tree decomposition is traversed in post-order,
and each tree decomposition node reflects a certain part of the problem instance.
During the traversal, one stores intermediate results in a table for each node.
Thereby, each node uses and transforms intermediate results 
of its child nodes and computes 
solutions to the corresponding problem part for the node. 
The dynamic programming algorithm relies on properties of the tree decomposition
in order to ensure that a non-empty table for the root node guarantees that
a problem solution was found.
Using this exact idea, one can also define algorithms to enumerate and count solutions.
%
%
%
The runtime of these dynamic programming algorithms
is polynomial in the instance size,
but additionally depends on a function $f(k)$ for some computable function~$f$,
where~$k$ is the parameter treewidth.
If the parameter~$k$ is reasonably small, such an algorithm
could outperform classical ASP solvers, 
which require exponential runtime in the instance size in the worst-case.
Since~$f$ might be exponential, we aim at classification of problems according to
the function~$f$ that is required to solve the problem using parameter treewidth.
These lower bounds are typically dependent on the widely believed \emph{exponential time hypothesis (ETH)},
which implies that there is no algorithm for \emph{Boolean satisfiability (SAT)}
such that satisfiability of a given formula with~$n$ variables can be decided in time~$2^{o(f(n))} \cdot n^{\mathcal{O}(1)}$.

\noindent\emph{Answer Set Programming (ASP).} 
ASP is a rule-based modeling and problem solving framework, where 
rules can contain (first-order) variables,
which are instantiated by the \emph{grounder}. 
The grounder is responsible for producing a \emph{(ground) ASP program}, which is a set of rules obtained by eliminating variables of a given \emph{non-ground} program.
%
%
Modern grounders parse non-ground programs that adhere to the \emph{ASP-Core-2}~\cite{aspcore2} syntax and output (among others) ground programs in \emph{SModels intermediate format}~\cite{aspcore2,GebserEtAl12}.
The main interest of this research proposal concerns the solving, i.e.,
how to efficiently evaluate an ASP program. 

%
\section{Research}%
First, we discuss the goals of the thesis.
Then, we provide a detailed description of achievements and the lessons learned so far.
Finally, we give an outlook about ongoing and future work.
\subsection{Goals}\label{sec:new}
The thesis shall address and achieve the following main goals.
\begin{itemize}
	\item Design ASP solving algorithms and techniques that utilize the structural parameter treewidth in order to efficiently solve ASP-Core-2 programs. 
	These algorithms shall be extended to problem variants and extensions later.
	\item Implement these ideas in a solver and compete against other ASP solvers and the related DynASP system.
	\item Investigate the theoretical limitations that any of these solvers can not evade given reasonable assumptions in computational complexity. The thereby obtained results might depend on certain fragments of ASP and could be crucial to further improve the solver.
	%
	%
	\item Generalize and apply these findings in a broader, general context. Especially applications in artificial intelligence might benefit from the outcome of this research. In this regard also other structural parameters could be considered.
	%
\end{itemize}%

\subsection{Achieved Results}\label{sec:app}
\label{sec:old}

First of all, we established dynamic programming algorithms~\cite{FichteEtAl17a} for decision and counting problems related to ASP\footnote{These algorithms were later also generalized to the formalism of default logic~\cite{FichteHecherSchindler18a}.}
using the full ASP-Core-2 syntax.
Thereby we adapted the existing DynASP solver~\cite{MorakEtAl11} resulting in the DynASP2 system.
This covers interoperation with state-of-the-art grounders, and handling of the SModels intermediate format including also optimization statements. 
The new system DynASP2
conceptually follows the approach of the original implementation,
where tree decomposition(s) of a certain graph representation of 
a given ground program are prepared\footnote{Later we also investigated the more general approach of fractional hypertree decompositions~\cite{FichteHecherLodhaSzeider18}, which could be relevant for ASP as well.}.
The resulting tree decomposition is then traversed
in a bottom-up manner (using post-order traversals) to evaluate the program. 
We implemented several variants of such algorithms,
based on the input program's graph representation.
One such graph representation is the 
\emph{primal graph representation}, where atoms are vertices, and atoms appearing together
in a rule have an edge between them.
Further, we also investigated algorithms based on the
\emph{incidence graph representation}, where atoms and rules are vertices, and when an atom appears in a rule there is an edge between them.
%
Then, we improved the whole approach, where we presented dynamic programming algorithms
that rely on multiple passes (where the tree decomposition is traversed multiple times).
This new approach of traversing in multiple passes, which extends an existing work on two passes~\cite{BliemEtAl16},
allows then to improve existing data structures of DynASP2.
In particular, one can benefit from early pruning of data after each traversal,
which then allows for efficient solving in practice~\cite{FichteEtAl17b}, given dedicated data structures 
that heavily rely on preventing redundancy using pointers.
Given ASP programs of small treewidth,
DynASP2 proved to be competitive in the setting of \emph{model counting}, 
a central problem in areas like
machine learning, statistics, probabilistic reasoning and combinatorics.
When counting answer sets, our approach conceptually has a big advantage: it does not require
to materialize the full answer sets in order to count them. This ensures huge
speed-ups against classical ASP systems like clasp~\cite{GebserEtAl12}. 
However, our implementation was also able to beat existing SAT model counters. 
Benchmarks indicate that the performance of model counting using a popular fragment of quantified boolean formulas with quantifier depth two
(2-QBFs) is competitive as well. 
%

Model counting proved to be a rather successful application
of dynamic programming on tree decompositions, since it 
lands itself well to parallelization. 
Indeed, using a novel approach on modern graphics computing units (GPUs),
we were able to compete existing model counters in a
comprehensive benchmark evaluation using all known state-of-the-art
systems for model counting of SAT formulas
and weighted variants. We recently improved the resulting system gpusat~\cite{FichteEtAl18a} by a new architecture involving data compression,
dedicated data structures, optimized counters and 
customized tree decompositions.

Later, we generalized our algorithms to \emph{projected model counting}, where models that are identical
with respect to a given set of projection variables, i.e., when ``projected'' to these projection variables,
count as one (projected) model. 
To this end, we designed an algorithm~\cite{FichteEtAl18} that works in multiple passes for SAT.
The idea was to propose one dynamic programming algorithm that relies on previous passes that
solve the base (decision) problem according to certain conditions. 
Then, this algorithm takes the results of the
previous passes and performs projected counting on top of it.
Unfortunately, the algorithm might exponentially increase (in the treewidth) the results of the previous passes in the worst case.
As an example, although SAT can be solved in single-exponential runtime in the treewidth (while still being polynomial in the instance size),
projected model counting on SAT requires double-exponential runtime in the treewidth using our algorithm.
However, we also considered the exponential time hypothesis (ETH), a standard assumption in computational complexity.
Our results reveal that if one assumes ETH, one cannot significantly improve the worst-case runtime of this problem.
We also generalized our results of projected model counting to abstract argumentation~\cite{FichteHecherMeier18c} and ASP~\cite{FichteHecher18a,FichteHecher19},
where we provided algorithms and lower bounds based on ETH for fragments of ASP (for decision, counting and projected model counting problems).
Some of these lower bounds were achieved by relying on our recent generalization of a result for 2-QBFs to arbitrary QBFs,
which provides a novel methodology for lower bounds of problems located higher in the polynomial hierarchy~\cite{FichteHecherPfandler19}.




\subsection{Current and Future Work}\label{sec:changes}
%
%
%
Given our recent dynamic programming algorithms~\cite{FichteHecher19} that utilize treewidth for certain fragments of ASP, we consider
an implementation, which will be added to DynASP. 
Especially in the context of \emph{hybrid parameterized solving}, which aims at the interplay
between existing monolithic solvers (e.g., clasp) and approaches based on exploiting (structural) parameters,
there is still room for improvement.
There are already existing systems of this kind, e.g., \emph{dynqbf}~\cite{CharwatWoltran19}, which is capable of deciding validity of QBFs 
up to treewidth 80 with this idea. Since DynASP can be used up to treewidth 14, and existing reductions
from ASP to QBF seem to fail for dynqbf, one might consider a new implementation DynASP3, which provides
the best of monolithic solving (e.g., clasp) and parameterized solving (e.g., DynASP2).

At the moment I am focusing on an efficient implementation of
our algorithm for projected model counting, since it seems that the mentioned worst-case result
does not reflect the average case.
In particular,
I am evaluating the prospect of a potential implementation targeted on alternative hardware, e.g., GPUs,
and potential strategies towards projected model counting solvers for ASP.
This new type of hardware raises interesting questions concerning 
alternative data structures and methods to exploit parallelism efficiently.

Further, we are permanently improving on our methodology~\cite{FichteHecherPfandler19} for lower bounds,
and the idea is to provide a catalog of simple, alternative proofs 
based on our methodology for existing lower bounds for treewidth under ETH.
It is for example still open, whether our methodology can be applied to easily show lower bounds 
that are called slightly superexponential~\cite{LokshtanovMarxSaurabh11} (in the treewidth), 
which is between single- and double-exponential.

\subsection*{References}

\renewcommand\refname{\vskip -0.98cm}
\footnotesize
\bibliographystyle{abbrv}\footnotesize
\bibliography{qbf_tw_lbs-cleaned}

\end{document}